# Assessing Hemodynamic Impact of Tissue-Engineered Vascular Graft Displacement: Combining Postoperative in vivo Results and Computational Modeling to Improve Surgical Planning

Seda Aslan[1,2], Enze Chen[3,4], Miya Mese-Jones[1,5], Jacqueline Contento[6], Hidenori Hayashi[7], Keigo Kawaji[8,9], Joey Huddle[10], Jed Johnson[10], Yue-Hin Loke[6], Mark Fuge[11,12], Laura Olivieri[6,13], Thao D Nguyen[1], Narutoshi Hibino[7], Axel Krieger[1]

[1]Department of Mechanical Engineering, Johns Hopkins University, Baltimore, MD, USA

[2]Department of Anesthesiology and Critical Care Medicine, Children's Hospital of Philadelphia, Philadelphia, PA, USA

[3]Department of Civil and Systems Engineering, Johns Hopkins University, Baltimore, MD, USA

[4]Department of Mechanical Engineering, University of Wisconsin–Madison, Madison, WI, USA

[5]Baltimore Polytechnic Institute, Baltimore, MD, USA

[6]Department of Cardiology, Children's National Hospital, Washington, DC, USA

[7]Department of Surgery, Section of Cardiac Surgery, University of Chicago Medical Center, Chicago, IL, USA

[8]Department of Biomedical Engineering, Illinois Institute of Technology, Chicago, IL, USA

[9]Department of Medicine, Section of Cardiology, University of Chicago Medical Center, Chicago, IL, USA

[10]Nanofiber Solutions, Dublin, OH, USA

[11]Department of Mechanical Engineering, University of Maryland, College Park, MD, USA

[12]Department of Mechanical and Process Engineering, ETH Zurich, Zurich, Switzerland

[13]Division of Pediatric Cardiology, University of Pittsburgh Medical Center, Pittsburgh, PA, USA

Corresponding authors: Seda Aslan, saslan2@jhu.edu, Axel Krieger, axel@jhu.edu

**Abstract**

**Purpose:** Tissue-engineered vascular grafts (TEVG) have shown promise in advancing vascular reconstructions. However, precise in vivo implantation is challenging, and it is unclear how deviations in location and size affect hemodynamics. This study aims to 1) compare preoperative designs and postoperative anatomies of TEVG in an in vivo study to evaluate discrepancies and 2) investigate the impact of graft displacement and size on hemodynamics by virtually simulating implantation scenarios that are informed by in vivo postoperative results.

**Methods:** Designed and postoperative geometries of four porcine aortas were compared to measure the mismatch in implantation location and graft shape. These results informed a virtual TEVG implantation study. TEVG location, orientation, and size were varied to investigate the effects on the final aorta shape and hemodynamics. Anastomosis of TEVG was simulated using finite element modeling. Key hemodynamic metrics were obtained from virtual implantations and actual postoperative anatomies using computational fluid dynamics.

**Results:** Our in vivo study showed that TEVGs can experience up to 6.9 mm displacement and a 38° rotational shift post-implantation, leading to discrepancies in pressure drop (2.5 mmHg, 50%) and time-averaged wall shear stress (7.2 Pa, 72%) compared to predictions. Virtual TEVG implantation showed that peak systolic pressure drop (PSPD) was most sensitive to translation in the inferior-superior direction and rotation about the anterior-posterior axis. Size mismatch had a greater impact on time-averaged wall shear stress (TAWSS) (85%) than PSPD (23%). Additionally, virtual anastomosis simulations improved aortic shape predictions by 27.5%.

**Conclusion:** Our results highlight the sensitivity of key hemodynamic metrics to graft implantation location and size mismatch. By quantifying displacement ranges and their impacts during surgery, surgeons can make informed decisions.

# 1  Introduction

Aortic anomalies, such as coarctation and arch hypoplasia, are congenital heart defects that affect 1-2% of newborns [1]. These conditions are characterized by narrowing of the aorta, the main artery carrying oxygenated blood from the heart to the body, which can impair blood flow and lead to serious complications [2]. Treatment typically involves transcatheter approaches, such as balloon angioplasty or stent placement [3, 4], or surgical interventions including resection with end-to-end anastomosis, interposition grafting, or patch aortoplasty [5, 6]. These procedures aim to enlarge the narrowed segment, restore proper hemodynamics, and reduce the risk of heart failure, stroke, or other organ damage [7].

Surgical intervention remains the first-line treatment for children diagnosed with complex aortic coarctation [8, 9]. The standard surgical approach typically involves using a patch or graft to reconstruct the aorta and restore normal blood flow [10, 11]. However, the use of grafts and patches presents challenges in cases of coarctation or arch hypoplasia due to the wide variability in anatomical presentation. Recently approved devices such as the GORE®TAG® and Thoraflex™ Hybrid provide effective repair options, yet carry notable limitations including risks of endoleak, rupture, migration, or kinking, and require long-term monitoring for durability [12–14]. Additionally, standardized grafts have limited applicability in patients with complex anatomy or extensive disease. They are also not ideal for pediatric patients, as they do not accommodate growth and typically necessitate additional surgeries over time.

Tissue engineering represents a promising approach for developing vascular grafts that serve as biodegradable scaffolds, gradually replaced by the patient's own tissue [15, 16]. By mimicking the mechanical properties of native vessels, tissue-engineered vascular grafts (TEVGs) have the potential to reduce complications and improve long-term outcomes. Studies in animal models have demonstrated their safety and efficacy in both short- and long-term applications [17, 18]. Patient-specific TEVGs have shown advantages over off-the-shelf grafts in complex reconstructions due to their superior anatomical conformity [17, 19–21].

While previous research demonstrated promising *in silico* [22] and *in vivo* [17, 19, 23] results validating the positive impact of patient-specific TEVG on hemodynamic performance, discrepancies have been observed between the planned and actual postoperative graft locations and overall aortic geometry. A center of gravity analysis [23] revealed that greater deviation between intended and observed TEVG position was associated with increased wall shear stress (WSS) in postoperative anatomies. Postoperative shape and hemodynamics predictions have been made possible through finite element modeling (FEM) and computational fluid dynamics (CFD) simulations [20, 22, 24–27]. For example, combined FEM and CFD simulations have laid the groundwork for surgical planning of arterial patch reconstructions by predicting postoperative shape and hemodynamics [28]. Multiscale simulations and virtual planning approaches via surgeon input have also been used to model reconstructed aortic arch geometry and blood flow [29]. Similarly, a study on virtual transcatheter stent-graft repairs in the aorta using FEM and CFD enabled direct comparison between preoperative predictions and postoperative outcomes in a single patient [30]. A more recent study using CFD has focused on the effects of TEVG oversizing and shown that inflow narrowing alters WSS and flow patterns, thereby influencing neo-vessel formation, and highlighted that placement of size-matched vascular conduits is critical for accommodating patient growth and material remodeling [31].

Despite these advancements, systematic comparison of preoperative predictions with in vivo postoperative outcomes of custom-designed grafts, particularly TEVGs, remains limited in larger patient cohorts. Furthermore, precise graft placement during surgery remains a clinical challenge, and the hemodynamic effects of placement inaccuracies have yet to be systematically studied. Defining acceptable tolerances for deviations in graft position, orientation, and size is crucial to understanding their impact on postoperative anatomy and flow dynamics.

To address these gaps, in this study, we report immediate postoperative outcomes of TEVG implantation in four porcine models (n=4) and investigate how variations in surgical precision affect postoperative hemodynamics. We hypothesize that incorporating soft tissue interactions into the virtual implantation framework will improve the accuracy of predicted outcomes. Our work integrates surgical planning through graft design, FEM, CFD simulations, and postoperative analysis. Beyond its methodological contributions, this work addresses a pressing clinical need: defining the tolerance for graft implantation deviations without compromising hemodynamics is critical for improving surgical reliability, especially in pediatric patients where anatomical complexity and long-term growth remain challenges. To our knowledge, this is the first systematic study to quantify the tolerance to graft misplacement, including deviations in location, orientation, and size, in the context of patient-specific TEVGs. Our findings provide a quantitative framework for evaluating the tolerance to graft implantation deviations and establish a foundation for future patient-specific surgical planning aimed at preserving preoperatively predicted hemodynamics.

## 2 Methods

Our study was divided into two parts: 1) an in vivo animal study to quantify the surgical discrepancies and 2) a virtual study to evaluate the hemodynamic effects of the deviations.

In the first part, an in vivo experiment was conducted to evaluate customized TEVGs, as previously described [17, 22]. The TEVGs were implanted in four porcine models (n=4), including three branched TEVGs (n=3) and one tubular TEVG (n=1), to assess the accuracy of surgical planning and implantation. Surgical planning involved computer-aided design (CAD) and CFD simulations to predict graft shape and hemodynamics before and after implantation. Postoperative TEVG geometries and hemodynamic data were then compared with preoperative predictions to quantify discrepancies.

In the second part, we created various TEVG implantation scenarios by virtual surgery. Data from the in vivo study informed these simulations, allowing us to model surgical imprecision and assess its impact on hemodynamics. The graft implantation simulations systematically varied the implantation location, TEVG orientation, and size to quantify the range of permissible imprecision where postoperative hemodynamics are not significantly affected. Soft tissue interactions between TEVG and native tissues were modeled using finite element methods (FEM) to mimic implantation. Details of the animal study and virtual simulations are described in the following subsections, and an overview of the study design is provided in Fig. 1.

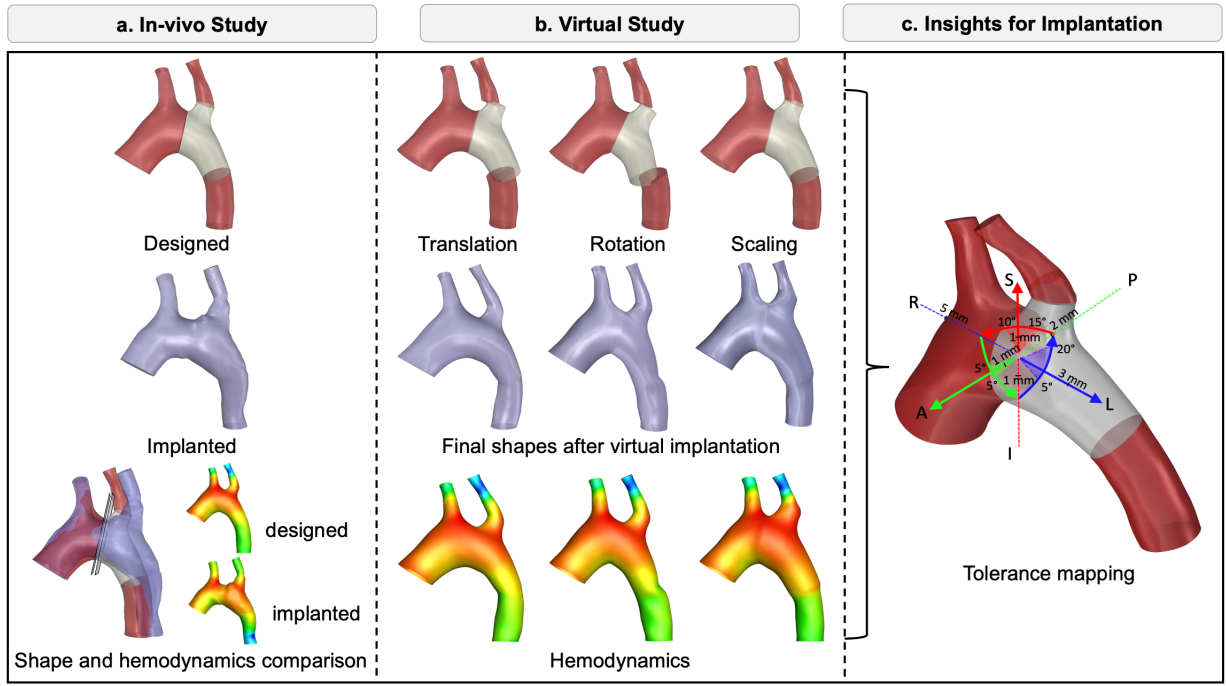

Figure 1: Overview of the study: *a.* In vivo study, including TEVG design, postoperative aortic shape acquisition, and comparison between designed and implanted geometries with corresponding hemodynamic analysis. *b.* Virtual study, incorporating variations in graft translation, rotation, and scaling based on in vivo results to evaluate the impact of implantation imprecision. *c.* Tolerance mapping, providing insights into acceptable deviations in TEVG location, orientation, and size that maintain hemodynamically safe outcomes.

## 2.1 In Vivo Study

### 2.1.1 Acquisition of Native Aorta Anatomy:

An Institutional Animal Care and Use Committee (IACUC)–approved study (#72605; approved 10/24/2019) was performed at the University of Chicago Medical Center. The heart anatomies of four porcine models (n=4, 20-30 kg) were obtained from cardiac magnetic resonance angiography (MRA). The details of image acquisition were explained in [17, 23]. MRA images were segmented using Mimics (Materialise, Leuven, Belgium) to identify the aorta. The resulting segmentation was then converted into a three-dimensional (3-D) anatomical model and smoothed.

### 2.1.2 Surgical Planning and Customized Graft Design:

After obtaining the 3-D geometry of porcine aortas, a customized graft design was performed. To account for growth during the 1-month period between imaging and implantation, the aorta model was expanded by 5% uniformly using the scaling tool in 3-matic (Materialise, Leuven, Belgium). This percentage was determined based on the growth curve of the mixed Yorkshire and Landrace porcine model [23]. The resulting scaled model was exported in stereolithography (STL) format.

TEVG design process for the aortic anatomy involved using CAD software (SolidWorks, Dassault Systèmes, Waltham, MA) to virtually resect and reconstruct the aorta. In surgery, a clamp is first placed on the aorta to create a temporary bloodless field, allowing resection of the stenotic region and reconstruction. To replicate this step, we simulated clamp placement between the two upper arch branches to define the region for TEVG design and implantation as shown in Fig. 2a. Clamp sizes (Beck Infant Aorta Clamp – Angled DeBakey Atraumatic Jaws, Wexler Surgical) were selected based on the aortic dimensions of each porcine model [22].

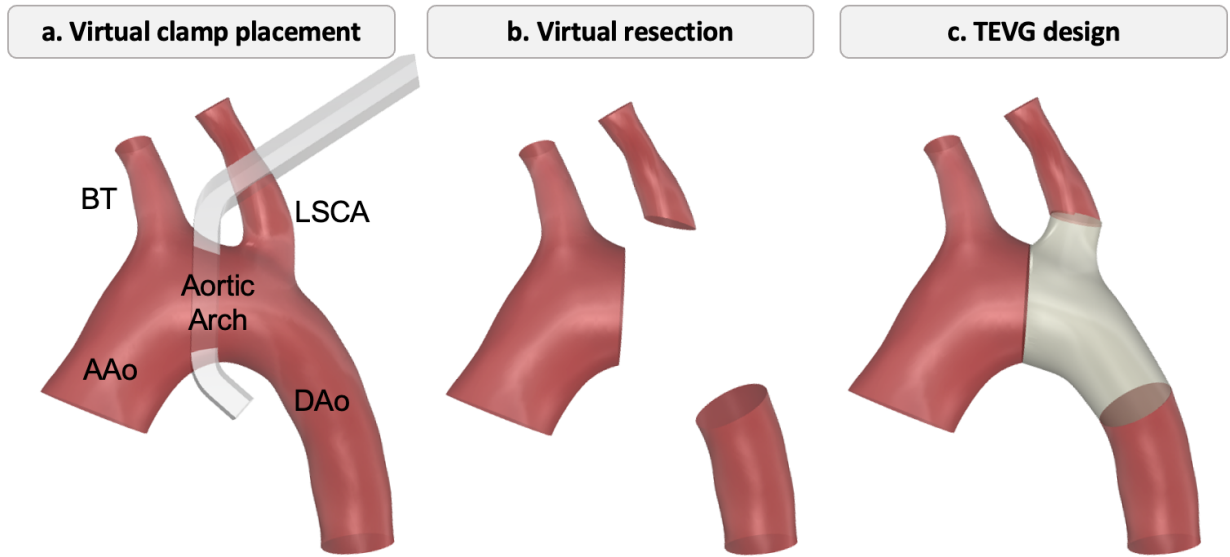

Figure 2: Design steps for customized TEVG. a. Virtual clamp placement between the brachiocephalic trunk (BT) and left subclavian artery, (LSCA) on the aortic arch to define the resection region. b. Virtual resection of the aortic segment. c. Reconstruction of the resected region using a patient-specific TEVG connecting the ascending aorta (AAo) to the descending aorta (DAo).

Subsequently, a virtual resection of the native arch region was executed, wherein the native aorta within the surgeon-specified locations was removed. This process isolated the area to be replaced by the graft as depicted in Fig. 2b. A customized graft that replaced the native shape was designed based on the native anatomy. The curvature and diameter of the aortic arch were varied to create either a tubular shape or a branched descending aorta graft. The design process followed the previous work [17, 19, 22]. Finally, the designed TEVG geometry shown in Fig. 2c (grey part of the aorta) was exported in STL format for manufacturing.

### 2.1.3 TEVG Fabrication and Implantation

The STL file of the designed graft was utilized for manufacturing a stainless-steel mandrel using 3-D printing through direct metal laser sintering (DMLS) at an external printing facility (Protolabs in Rosemount, MN or Xometry in Gaithersburg, MD). A biodegradable nanofiber material, comprising a 1:1 ratio of poly-caprolactone (PCL) and poly-L-lactide-co-ε-caprolactone (PLCL), was coated onto the steel mandrel at Nanofiber Solutions (Dublin, Ohio) through the electrospinning process, maintaining a high voltage during fiber application. The resulting nanofiber scaffold was carefully detached from the metal mandrel, and the biodegradable graft was then placed inside a standard Tyvek pouch. For sterilization, the graft underwent a low-temperature process utilizing vaporized hydrogen peroxide and ozone with the STERIZONE VP4 system from Getinge, Sweden [23].

TEVG implantation surgery was conducted through a left thoracotomy under general anesthesia. Aortic grafts were implanted either in the descending aorta (DAo) (n = 1) or both in the DAo and the left subclavian artery (LSCA) (n = 3). For aortic grafts involving the LSCA (branch), the anastomosis was performed distal to the brachiocephalic trunk (BT), whereas grafts without a branch were positioned distal to the left LSCA. During this procedure, a partial bypass was employed [17]. TEVGs were explanted at the acute phase, within 2 days following the surgery.

### 2.1.4 Postoperative Analysis

The aorta images with implanted TEVG were obtained post-surgery, before explantation. Postoperative and designed aorta geometries were aligned for comparison. First, the proximal regions of pre-operative and postoperative aortas including AAo and BT were aligned by transforming postoperative geometry onto the same region of pre-operative geometry, as shown in Fig. 3a. Then, two virtual anastomosis planes were defined on the designed and postoperative geometries based on where the graft was planned to be implanted and the location where the surgeon implanted the graft in vivo. These planes were used to measure the discrepancy in implantation location. The transformation between the designed and implanted TEVG was calculated from rigid registration using 3D Slicer [32]. The designed and postoperative anastomosis planes are shown in Fig. 3a.

On the designed anastomosis plane, at the center of the graft, a new anatomical coordinate system was created as shown in Fig. 3b to calculate the displacements of the TEVG in left-right (L-R), inferior-superior (I-S), and anterior-posterior (A-P) directions, as well as the rotation angles about each axis. The volumes of the designed and postoperative graft shapes were also obtained to calculate the graft size mismatch.

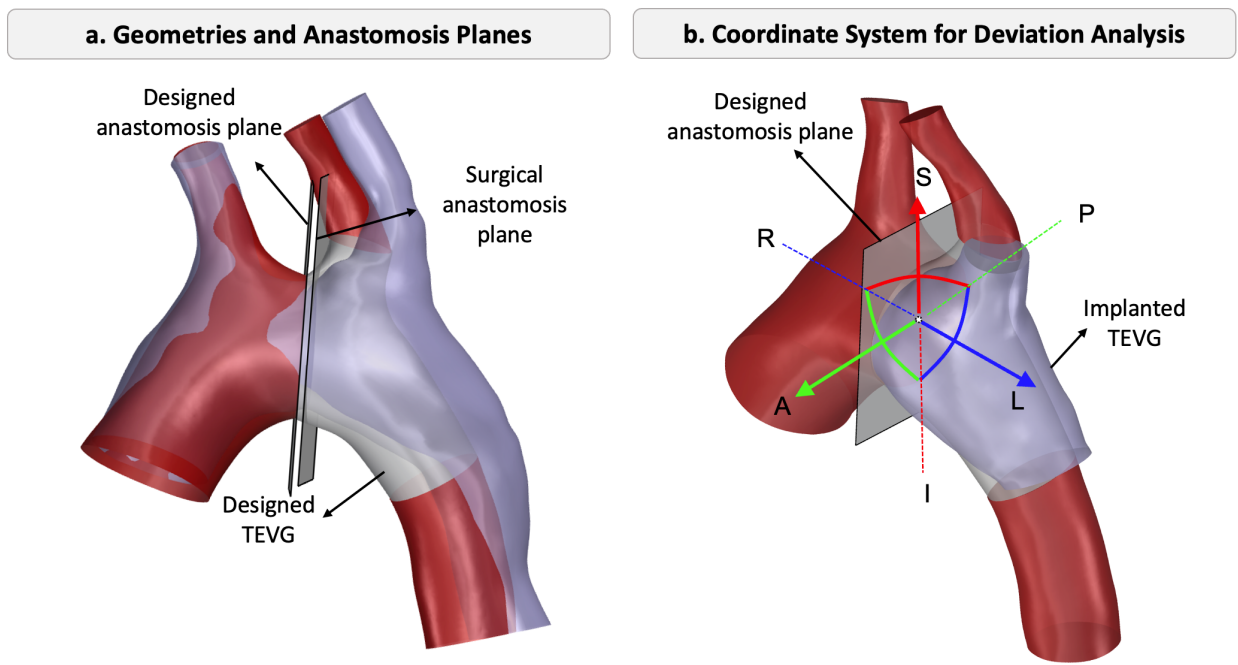

Figure 3: a. Alignment of designed and postoperative aortic geometries, showing the designed and surgical anastomosis planes. b. The isometric view illustrates the implanted TEVG, designed anastomosis plane, and defined anatomical directions (A: anterior, P: posterior, R: right, L: left, S: superior, I: inferior).

## 2.2 Virtual Study

The second part of this work focused on simulating TEVG implantation informed by the discrepancies observed in the in vivo study. We examined the impact of surgical imprecision and offered insights to enhance the accuracy of postoperative shape and hemodynamic predictions. The virtual implantation simulations modeled soft tissue deformation resulting from TEVG implantation in the native aorta, creating various scenarios to investigate graft displacement and size mismatches and their individual effects on hemodynamics.

Graft displacement arises from deviations in the location of the native aorta incision during the implantation procedure. Achieving a precise cut at the intended location during surgery is inherently challenging. Variations in the anastomosis location may occur, resulting in the translation or rotation of the TEVG relative to its intended position, which can significantly influence the aorta's shape and hemodynamics.

In this part, we performed virtual anastomosis for both a tubular and a branched aorta graft. For the tubular TEVG, anastomosis was performed distal to the aortic arch. For the branched TEVG, the anastomosis was placed between two branches, BT and LSCA. After virtual anastomosis plane was created at the proximal side where the graft was sutured to the native tissue, a new coordinate system was established, with axes aligned in the I-S, A-P, and L-R directions, (Fig. 3b). The translation and rotation thresholds were derived from in vivo results by calculating the mean and adding two standard deviations, ensuring coverage of the possible variations. For L–R translations, both the TEVG and the anastomosis plane were translated, resulting in a shift of the resection location. In contrast, A–P and I–S translations were applied only to the TEVG, as translating the plane in these directions would not alter the resection site. Similarly, the rotation was applied to both TEVG and anastomosis plane where TEVG is connected to distal AAo (or proximal DAo in tubular TEVG case).

Another source of discrepancy between the designed and postoperative geometry is the size mismatch between the designed and manufactured TEVGs. To investigate the impact of size variation, we scaled the TEVGs uniformly. Potential scenarios for displacement and size variation are illustrated in Fig. 4.

By incrementally varying the implantation location, graft orientation, and size, this study evaluated the sensitivity of hemodynamics and final shape to elucidate the importance of implantation precision and quantify the permissible range of imprecision without significantly compromising hemodynamic outcomes.

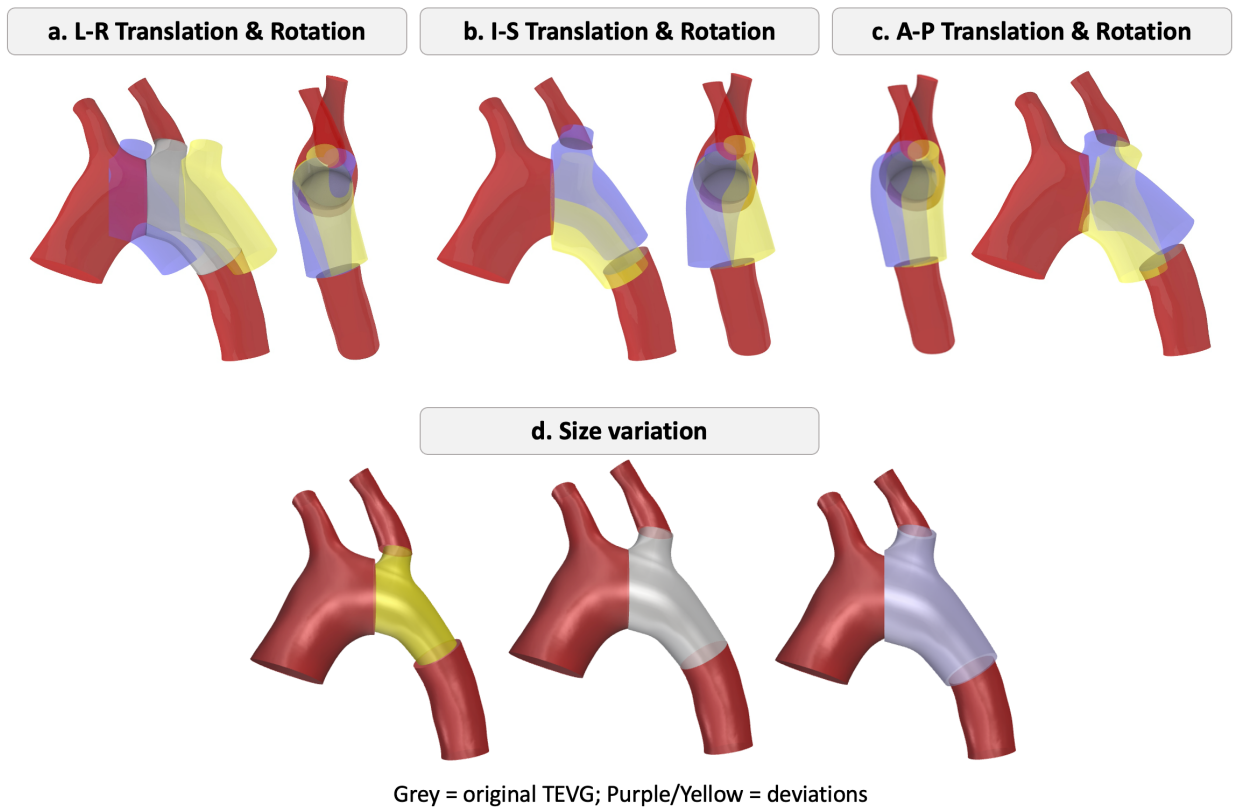

Figure 4: Virtual implantation scenarios illustrating translational, rotational, and size deviations of the TEVG relative to the designed baseline (grey). (a–c) Translation and rotation in left–right (L–R), inferior–superior (I–S), and anterior–posterior (A–P) directions, respectively. (d) Variation in graft size. Purple and yellow indicate deviations in opposite directions: (a) *Right* (purple) vs. *Left* (yellow); (b) Superior (purple) vs. Inferior (yellow); (c) Anterior (purple) vs. Posterior (yellow). For size variations (d), purple denotes a larger graft and yellow denotes a smaller graft.

### 2.2.1 Simulations of TEVG implantation

The surgical implantation of TEVG involves resecting the native aorta tissue and replacing it with the designed graft by performing an end-to-end anastomosis. To simulate the surgical implantation of TEVG, we utilized FEM. The FE model consisted of four parts: the native ascending aorta (AAo), descending aorta (DAo), an upper branch, and the tubular TEVG, as shown in Fig. 5. The walls of both the native tissues and the TEVG were represented as shell structures and discretized with quadrilateral elements in Coreform-Cubit (Orem, Utah).

At each anastomosis location, the graft edge and the corresponding native vessel edge (AAo, DAo, and branch) were meshed with an equal number of nodes to facilitate node-to-node pairing, representing surgical suturing. For each graft edge node (source), the corresponding native tissue node (target) was identified by comparing directional vectors and minimizing the relative angular difference, ensuring that nodes with the closest spatial orientation were paired. This procedure was repeated for all anastomotic interfaces, resulting in consistent and anatomically realistic node alignments across regions.

After pairing the nodes, displacement boundary conditions were prescribed to complete the connection (i.e. anastomosis). To enforce the displacement boundary conditions, auxiliary dummy nodes were created for each paired node at the anastomosis interface. The distance between paired nodes was first calculated, and relative displacements from the dummy nodes were used to prescribe the coupling constraints. This was implemented through multi-point constraints:

$$\boldsymbol{u}_{asc}^i - \boldsymbol{u}_{TEVG}^i - \boldsymbol{u}_{dummy}^i = 0, \qquad u_{dummy}^i = -(\boldsymbol{X}_{asc}^i - \boldsymbol{X}_{TEVG}^i)$$

where $\boldsymbol{u}_{asc}^i$ and $\boldsymbol{u}_{TEVG}^i$ are the displacements of the corresponding native and graft nodes, respectively, and $\boldsymbol{X}_{asc}^i$ and $\boldsymbol{X}_{TEVG}^i$ are their initial coordinates. Substituting yields

$$\boldsymbol{u}_{asc}^i = \boldsymbol{u}_{TEVG}^i - (\boldsymbol{X}_{asc}^i - \boldsymbol{X}_{TEVG}^i)$$

This formulation allowed the paired nodes to coincide after deformation. Finally, fixed boundary conditions were applied at the proximal edge of the AAo and the distal edges of the branches and DAo to mimic the in vivo tethering of the aorta to surrounding tissues.

FE simulations were performed using Abaqus 2019 software (Dassault Systemes, Vélizy-Villacoublay, France). The deformations of native aorta tissues and the TEVG were modeled using the hyperelastic Yeoh constitutive model [33]. For modeling the native aorta walls, we adapted the material constants from a previous study that measured the biomechanical properties of ex-vivo porcine aorta tissues and fitted the parameters to the Yeoh model [33]. We modeled the TEVG walls to be two and a half times less compliant than the native walls, based on experimental studies of porcine TEVGs explanted after one month of implantation [17].

The strain energy density function is given by:

$$W = C_{10}(I_1 - 3) + C_{20}(I_1 - 3)^2 + C_{30}(I_1 - 3)^3$$

where $I_1$ is the first invariant of the right Cauchy–Green deformation tensor, and $C_{10}$, $C_{20}$, and $C_{30}$ are material parameters obtained from experimental fitting.

Upon convergence of the simulations, the final shape of the deformed aorta was obtained. A video of TEVG implantation simulation and the final shape of the aorta were provided in Supplementary Material.

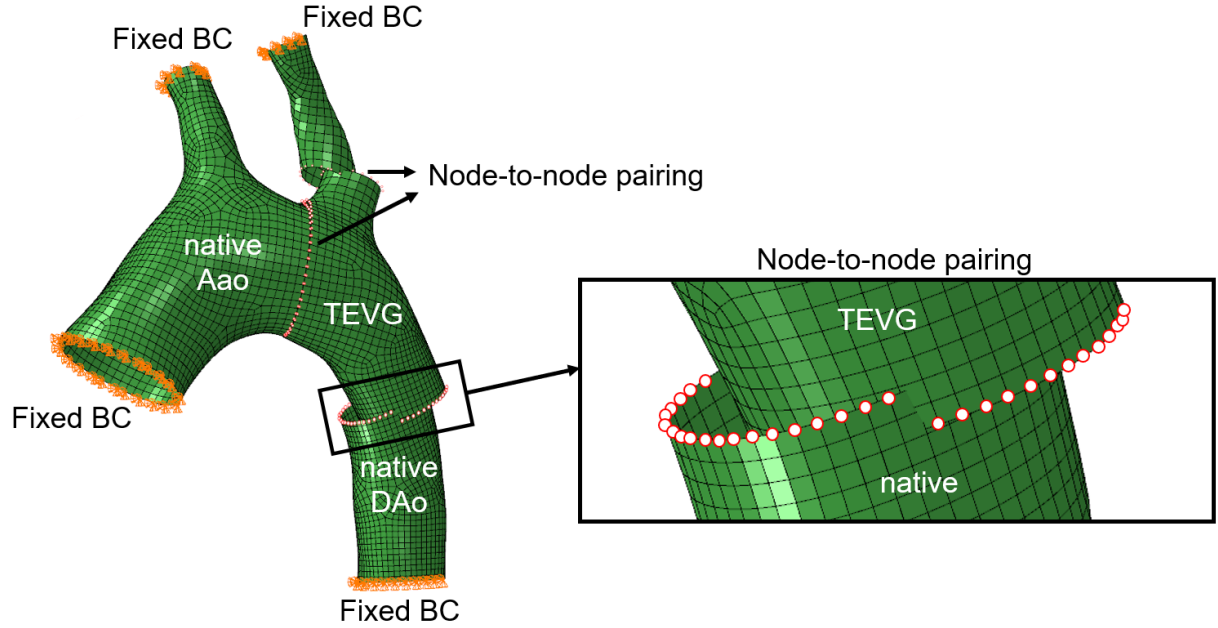

Figure 5: Computational model with boundary conditions and node-to-node pairing. Fixed boundary conditions (BCs) were applied at the proximal ascending aorta (AAo) and distal descending aorta (DAo). The inset highlights the nodes of the native tissue and TEVG that were paired to define the anastomosis.

### 2.2.2 CFD Simulations

Following the FE simulations, the deformed aortic geometries were discretized using tetrahedral elements with an inflation layer, as detailed in prior work [20]. Ansys Fluent was used to solve 3-D continuity and Navier-Stokes equations:

$$\frac{\partial \rho}{\partial t} + \nabla(\rho \vec{u}) = 0$$

$$\rho \frac{D\vec{u}}{D\vec{t}} = -\nabla p + \rho \vec{g} + \nabla \tau_{ij}$$

where $\rho$ is density, $t$ is time, $\vec{u}$ is velocity, $p$ is pressure, $\vec{g}$ is gravity, and $\tau_{ij}$ is stress tensor. Blood was modeled as non-Newtonian fluid with a density of 1060 kg/m³ and viscosity of 0.00371 Pa·s [20]. Details of the simulation setup are provided in previous work [22]. Briefly, the native and deformed aortic walls were assumed to be rigid to reduce computational cost, as prior studies demonstrated that the rigid wall assumption introduces negligible differences compared to compliant wall fluid–structure interaction models [34]. A time-dependent flow rate was prescribed at the inlet, while Windkessel models were applied at the outlets to capture downstream resistances and compliances [21]. Turbulent flow was modeled using the $k$-$\varepsilon$ model until convergence was achieved, over 5-10 cardiac cycles.

## 3 Results

The results are presented in two parts. First, the designed aortic geometries were compared with in vivo postoperative outcomes, and graft displacement and size mismatch were quantified. In the second part, results from the virtual TEVG implantation study are presented, where the effects of graft location, orientation, and size on hemodynamics, specifically PSPD and TAWSS, were investigated.

### 3.1 In vivo Study

The postoperative aortic geometries were superimposed onto the designed geometries, as shown in Fig. 6. In the overlays, red represents the native aorta, while the designed and postoperative TEVGs are shown in grey and dark purple, respectively. Light purple denotes the postoperative aorta. In the postoperative scans, we observed that in porcine 1, the second supra-aortic branch (LSCA) was absent, and in porcine 4, a stenosis developed in the mid-arch region following TEVG implantation.

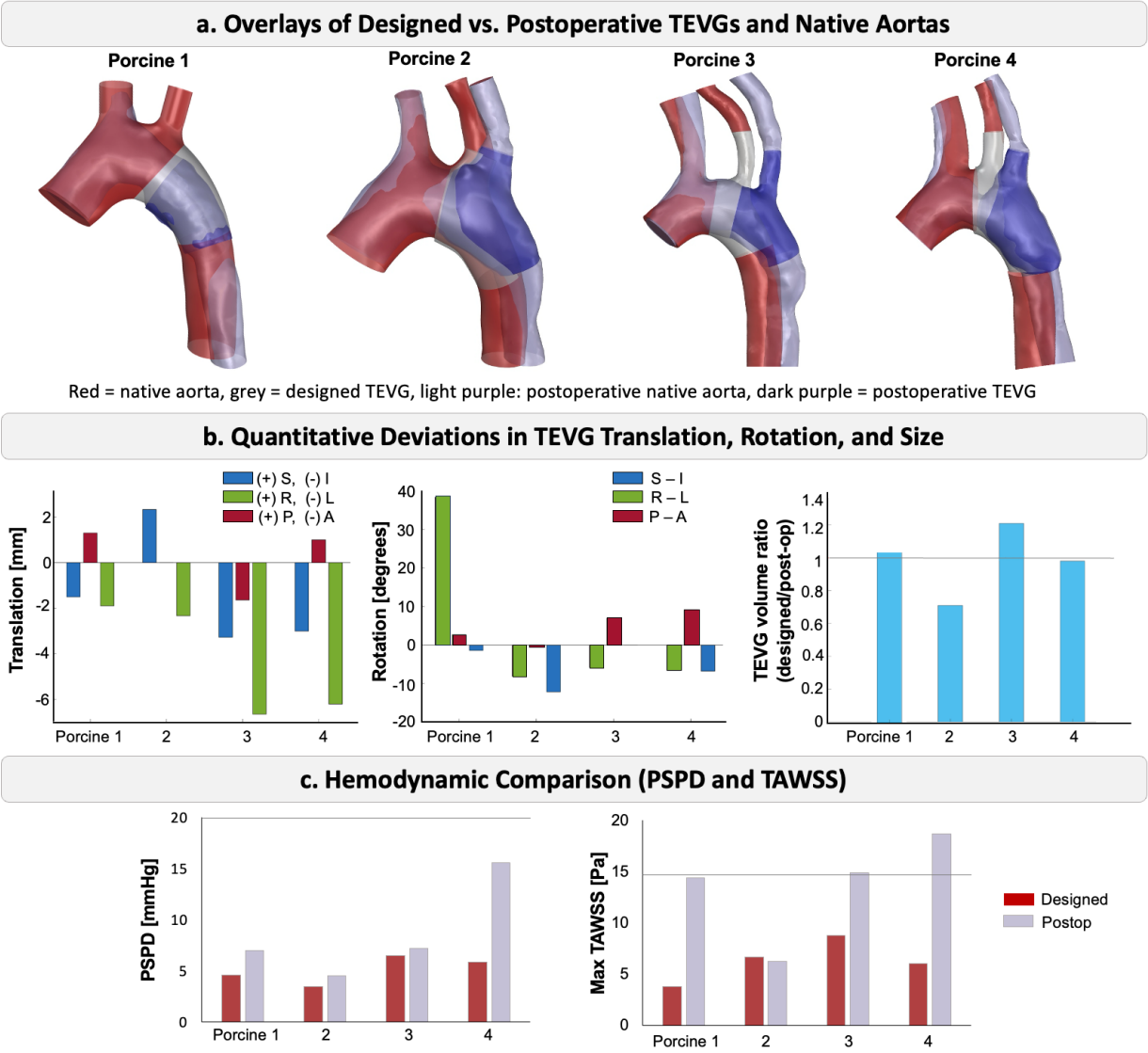

Figure 6: Comparative analysis of designed versus postoperative TEVGs. a. Overlays of designed and postoperative TEVGs and native aorta (red) for each porcine model. b. Quantitative deviations between designed and postoperative TEVGs: translations (mm), rotations (°), and graft volume ratio (designed/post-op). Positive/negative values indicate displacement or rotation directions. c. Hemodynamic comparison (PSPD and max TAWSS) of the designed and postoperative models.

The translation of the TEVGs from the intended implantation location was measured. Anatomical left displacement of the TEVG was observed in all animals, with the largest displacement being 6.9 mm in porcine 3. Inferior (I) displacement occurred in three out of four porcine, while displacement in the P-A direction was minimal compared to other directions. Rotational deviations ranging from 0 to 12° were observed in the branched TEVGs. The tubular TEVG in porcine 1 experienced a 38° rotation about the L-R axis. The presence of a branch in the TEVG aided in achieving a more precise rotational alignment during implantation.

Comparative analysis between the designed and postoperative geometries indicated that the size of the TEVG differed from the intended design. For all four implanted TEVGs, the ratio of designed graft volume to postoperative graft volume ranged from 0.7 to 1.2, with the designed graft being larger than the implanted one in two of the cases.

Deviations in TEVG location, shape, and size resulted in hemodynamic discrepancies between the designed and postoperative models. We evaluated PSPD and TAWSS, two clinically relevant metrics. PSPD between the AAo and DAo was consistently higher in postoperative geometries compared to the designed models. Similarly, maximum TAWSS values were greater in three of the four postoperative aortas, with porcine 2 being the exception due to its larger postoperative graft size. Notably, TAWSS was significantly elevated in porcine 1, 3, and 4, and porcine 4 also exhibited a pronounced PSPD increase caused by the stenosis. These findings align with the observed geometric deviations: porcine 3 and 4 had the largest displacements, and porcine 1 exhibited a significant size and rotational deviation. Collectively, the results emphasized that precise implantation is critical, as deviations can undermine the intended hemodynamic benefits of patient-specific TEVGs.

## 3.2 Virtual Study:

Based on the postoperative displacements of the TEVGs and discrepancies in overall aortic shape, we created virtual implantation scenarios by systematically rotating and translating the anastomosis plane in all directions and scaling the TEVG size. For each translation direction, rotation axis, and size mismatch, the mean and ±2 standard deviations derived from the in vivo study were used to establish variation thresholds in graft location, orientation, and size and to assess their individual and relative effects on hemodynamics.

To evaluate the sensitivity of hemodynamics to implantation deviations, translations were applied in 1-mm increments along each direction, and rotations were applied in 5° increments about each axis, and the graft size was varied in 5% increments. Using the designed location, orientation, and size as the benchmark, percentage differences in PSPD and maximum TAWSS were calculated.

The results of the virtual study are summarized in Fig. 7, where sensitivity to implantation deviations is expressed as percentage differences in PSPD and TAWSS. In Fig. 7a–b (left panels), line plots show PSPD and TAWSS differences as functions of translational displacement (mm) and rotational deviation (°). These results indicate that both PSPD and TAWSS were most sensitive to translations along the I–S direction, while translations in the L–R and A–P directions had smaller effects. For rotations, in terms of percent differences, the PSPD was relatively insensitive across all axes, whereas TAWSS showed greater sensitivity, particularly for rotations about the A–P and L-R axes.

The 2D heatmaps in Fig. 7a–b (middle panels) capture combined directional effects. These maps confirm that deviations in I–S translation and A–P rotation produced the largest increases in PSPD and TAWSS, while other directions and axes contributed less prominently.

Fig. 7c illustrates the effect of TEVG size variation. Undersized grafts (smaller than the design) produced larger deviations in both PSPD and TAWSS compared to oversized grafts, highlighting the importance of preventing undersizing during implantation.

The tolerance mapping in Fig. 7d illustrates the translational and rotational thresholds within which the TEVG could be positioned while maintaining TAWSS and PSPD within physiologic limits across all porcine aortas. Hemodynamically safe limits were defined as a PSPD < 20 mmHg and TAWSS between 0.4 and 15 Pa. A PSPD exceeding 20 mmHg indicates a large pressure drop across the repair site, commonly used clinically as a marker of stenosis and the need for reintervention [35], whereas TAWSS values above 15 Pa are associated with thrombosis risk and values below 0.4 Pa suggest susceptibility to atherosclerosis [36]. Across all simulations, PSPD remained below the 20 mmHg threshold, indicating that TAWSS served as the more restrictive criterion. The tolerance mapping therefore delineates the allowable implantation deviations that preserved both PSPD and TAWSS within physiologic limits. For the aortic geometries analyzed, the allowable translational deviations were approximately 3 mm in L, 5 mm in R, 1 mm in A, I, and S, and 2 mm in P directions. The corresponding rotational tolerances, defined according to the right-hand rule, were approximately 5° about the R, A, and P axes, 10° about S, 15° about I, and 20° about L.

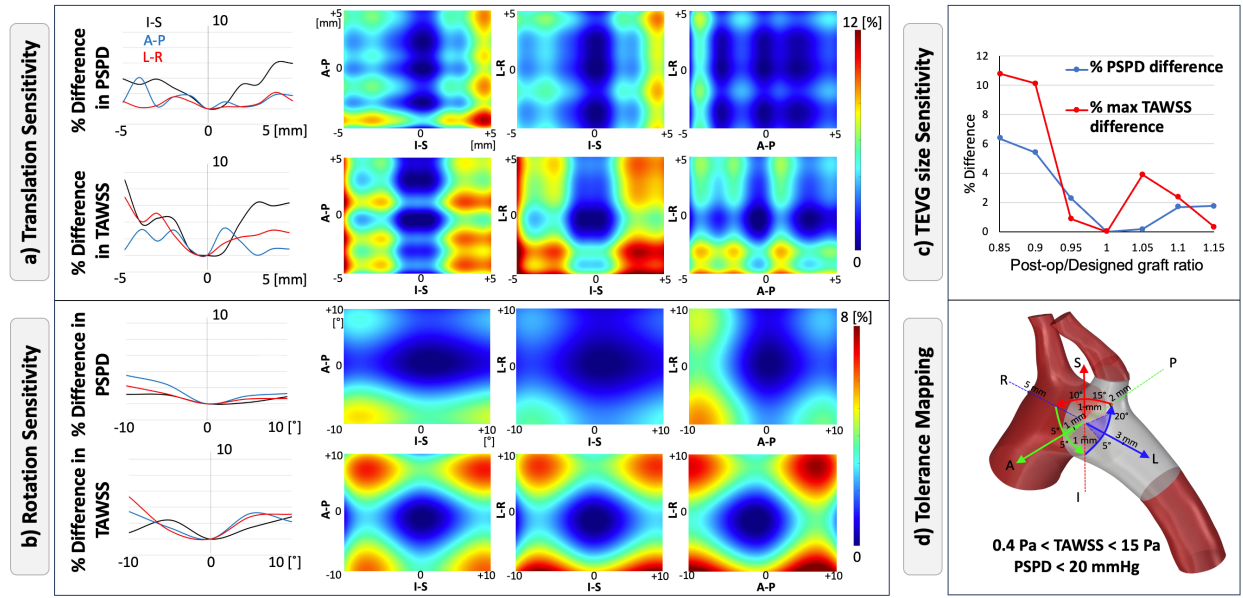

Figure 7: Sensitivity of PSPD and TAWSS to implantation deviations. (a) Translation: percentage differences in PSPD and maximum TAWSS as a function of displacement in I–S, A–P, and L–R directions (line plots, left) and combined directional effects (heatmaps, right). (b) Rotation: percentage differences in PSPD and maximum TAWSS as a function of angular deviation about I–S, A–P, and L–R axes. (c) Size: percentage differences in PSPD and TAWSS across TEVG size ratios (designed/post-op). (d) Tolerance mapping: anatomical reference showing the range of translations, rotations, and size mismatches that maintained hemodynamics within clinically safe limits (0.4 Pa < TAWSS < 15 Pa, PSPD < 20 mmHg).

To assess the accuracy of the virtual anastomosis simulations in reproducing postoperative aortic geometries, the rigid-body transformations (translation, rotation, and scaling) observed in vivo were applied to the original TEVG designs. The resulting FE-predicted geometries were then superimposed onto the corresponding postoperative aortic shapes in Fig. 8. In Fig. 8a, the original aortic designs were compared to the postoperative shapes, while in Fig. 8b, the transformed and deformed TEVG geometries are compared to the same postoperative shapes. Hausdorff distances [37] were computed to quantify surface distances between the two aortic geometries. The average surface distance between the postoperative aorta shapes and the original designs was 5.8 mm. The virtual anastomosis using FE simulations reduced this discrepancy by 27%, bringing the average mismatch down to 4.4 mm.

Despite this improvement, residual discrepancies remained after applying the in vivo transformations and re-simulating implantation (Fig. 8b). These differences were partly attributable to substantial postoperative geometric changes captured on imaging. Specifically, arch stenosis developed in porcine 4, and the second supra-aortic branch was not captured in the in vivo imaging of porcine 1, limiting geometric correspondence and reducing predictive accuracy.

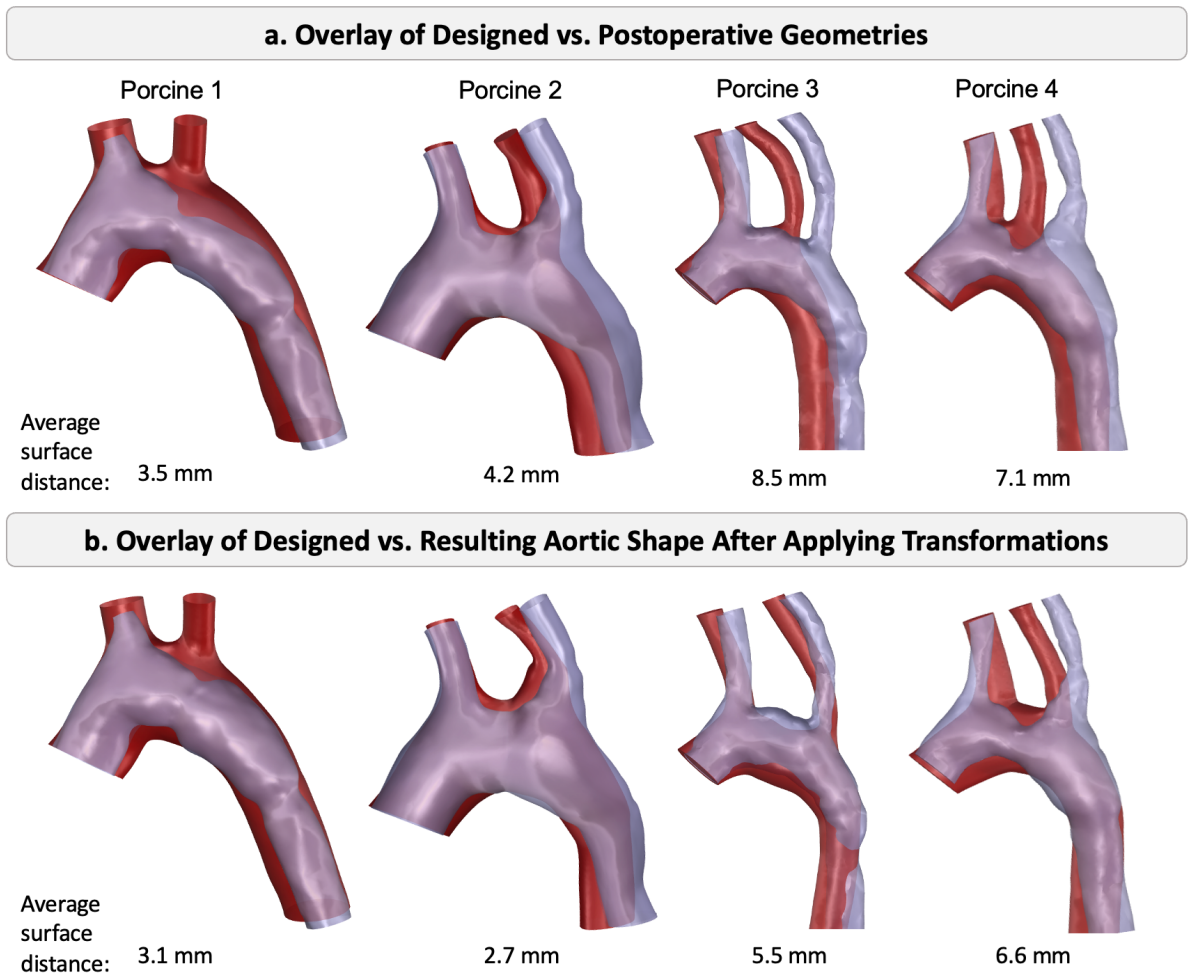

Figure 8: Comparison of aortic shapes. (a) Original aortic design (red) versus postoperative aortic shape (purple). (b) Resulting aortic shape (red) after applying the postoperative transformations and simulating the anastomosis of the designed TEVG. The numbers below each image represent the average surface distance between the two aortic shapes.

## 4  Discussion

This study investigated postoperative geometric changes following TEVG implantation and examined how deviations in graft position, orientation, and size influence postoperative hemodynamics. By integrating in vivo imaging from porcine models with FE and CFD simulations, this work builds on recent advances in customized graft modeling for aortic and congenital heart repair. Prior studies have demonstrated the successful design and implantation of patient-specific TEVGs and their potential to improve hemodynamic outcomes [17–19, 22]. In the present work, we extend these efforts by quantifying the impact of surgical placement accuracy on postoperative predictions, providing directional tolerance ranges for TEVG implantation.

Computational modeling has been instrumental in evaluating hemodynamics of cardiovascular reconstructions. Established workflows coupling imaging, virtual reconstruction, and hemodynamic simulation have enabled prediction of postoperative performance [20–22, 24–26, 29]. When applied to congenital repairs, particularly Fontan pathways, simulation-to-clinical comparisons have shown general consistency yet highlighted unavoidable geometric discrepancy. Haggerty et al. [38] demonstrated that small intraoperative differences can lead to deviations from preoperative simulations. Yang et al. [39] identified anastomosis location as a key determinant of hepatic flow, and Trusty et al. [40] showed that graft insertion location correlated with postoperative hemodynamic changes. Together, these studies underscored that even slight deviations in geometry can alter flow patterns.

The first analysis of TEVG implantation deviations in porcine RVPA and aortic models using pre- and postoperative imaging showed that offset in graft placement increased WSS [23]. Accurate virtual implantation also requires capturing soft-tissue interaction between grafts and native tissue, a principle demonstrated in FE-based simulations of stent expansion [24], stent-graft repair [30], and patch implantation [25, 28]. Building on these frameworks, the present work integrates deformation informed by pre- and postoperative anatomy to evaluate how TEVG placement variability affects postoperative PSPD and TAWSS. This approach adds a quantitative tolerance-evaluation component to the existing computational planning, linking geometric precision to hemodynamic consequences.

Our study is subject to several limitations. The short-term postoperative follow-up captured only early geometric outcomes; serial imaging will be needed to characterize longer-term remodeling, which is known to occur in TEVGs [17, 18], and relate early placement accuracy to subsequent adaptation. The FSI was not modeled, which may underestimate deformation under physiological pressures. Future simulations will incorporate FSI and measured or estimated pressure loading. Postoperative graft dimensions deviated from the original design likely due to fabrication variability and intraoperative pressure-induced changes. Improving manufacturing consistency and embedding physiological pressure loading in the modeling framework will enhance prediction fidelity. The aorta and TEVG were modeled as shell structures. However, native vessels and engineered grafts have layered, anisotropic behavior. Incorporating multilayer wall models and material anisotropy would better represent biomechanics. Finally, limited imaging field of view constrained boundary condition placement and contributed to branch alignment differences, larger field-of-view imaging will reduce this uncertainty.

## 5  Conclusion

This work introduces a quantitative framework for evaluating implantation tolerances in TEVG-based aortic repair. By integrating in vivo imaging with FE and CFD simulations, we link geometric precision to hemodynamic performance and present tolerance mapping that can be directly applied to personalized presurgical planning and intraoperative guidance for surgeons. These findings support the use of computational modeling to inform preoperative planning and guide manufacturing and surgical strategies for patient-specific TEVGs. Future studies will extend this framework to longer-term remodeling, larger cohorts, and intraoperative guidance integration to support clinical translation.

# Declarations

**Ethics approval and consent to participate**

The use of animal data was approved by the Institutional Animal Care and Use Committee (IACUC) (#72605, approved 10/24/2019). Consent to participate is not applicable. This study did not involve human participants. Clinical trial number: not applicable.

**Consent for publication**

Not applicable. This study did not involve human participants.

**Availability of data and material**

The datasets are available from the corresponding author on reasonable request.

**Competing interests**

J. Johnson, A. Krieger, and N. Hibino are inventors listed on International Patent WO/2017/035500Al (Patient-Specific Tissue Engineered Vascular Graft Utilizing Electrospinning). J. Johnson is an equity holder in Nanofiber Solutions. A. Krieger and is founder of and holds Shares of Stock Options in CorFix Medical, Inc. The results of the study discussed in this publication could affect the value of CorFix Medical, Inc. This arrangement has been reviewed and approved by Johns Hopkins University in accordance with its conflict-of-interest policies. All other authors have nothing to disclose regarding commercial support.

**Funding**

This project is supported by National Institutes of Health award numbers K25HL141634, R01HL143468, R21HD090671, and R33HD090671. The content is solely the responsibility of the authors and does not represent the official views of the National Institutes of Health.

**Authors' contributions**

**Seda Aslan:** Writing - original draft, Methodology, Investigation, Formal analysis, Visualization, Conceptualization. **Enze Chen:** Writing - review & editing, Methodology. **Miya Mese-Jones:** Writing - review & editing, Formal analysis. **Jacqueline Contento:** Writing - review & editing. **Hidenori Hayashi:** Writing - review & editing. **Keigo Kawaji:** Writing - review & editing. **Joey Huddle:** Writing - review & editing. **Jed Johnson:** Writing - review & editing. **Yue-Hin Loke:** Writing - review & editing. **Mark Fuge:** Writing - review & editing, Funding acquisition. **Laura Olivieri:** Writing - review & editing. **Thao D Nguyen:** Writing - review & editing, Supervision. **Narutoshi Hibino:** Writing - review & editing, Conceptualization, Resources, Funding Acquisition. **Axel Krieger:** Writing - review & editing, Conceptualization, Supervision, Funding Acquisition, Resources, Project Administration.

**Acknowledgments**

The authors acknowledge the supercomputing resources at Johns Hopkins University (https://www.arch.jhu.edu) which were essential for conducting the research reported in this paper.

Authors also thank the University of Chicago Animal Resources Center (RRID: SCR_021806).